\documentclass[journal]{IEEEtran}
\usepackage{latexsym}
\usepackage{amsfonts}
\usepackage{amssymb}
\usepackage{amsbsy}
\usepackage{amsmath}
\usepackage{amsthm}
\usepackage{color}
\usepackage{pgf}
\usepackage{graphicx,lscape,rotating,subfigure}
\usepackage{balance}
\usepackage[varg]{txfonts}
\usepackage[belowskip=-6pt,aboveskip=0pt]{caption}
\usepackage{algorithm,algpseudocode}
\usepackage[noadjust]{cite}

\theoremstyle{remark}

\theoremstyle{definition}

\title{\LARGE Reliable Uplink Communication through Double Association\\ in Wireless Heterogeneous Networks}

\author{Dong~Min~Kim,~\IEEEmembership{Member, IEEE} and Petar~Popovski,~\IEEEmembership{Fellow, IEEE}%
\IEEEcompsocitemizethanks{\IEEEcompsocthanksitem
The authors are with the Department of Electronic Systems, Aalborg University, Denmark (email: \{dmk;petarp\}@es.aau.dk).
}%
\thanks{This work has been supported by the Danish High Technology Foundation via the Virtuoso project.}
}

\begin{document}
\maketitle

\begin{abstract}
We investigate methods for network association that improve the reliability of uplink
transmissions in dense wireless heterogeneous networks. The stochastic geometry
analysis shows that the double association, in which an uplink transmission is
transmitted to a macro Base Station (BS) and small BS, significantly improves the
probability of successful transmission.
\end{abstract}

\section{Introduction}

Traditionally, the focus of wireless cellular networks has been on the downlink (DL)
traffic due to its higher volume compared to the uplink (UL) traffic. Recent developments
show a clear trend in the increase of the UL traffic due to new mobile applications, such
as social networks, cloud backup storage, video chatting etc., as well as the explosive
growth of machine-to-machine (M2M) connections. M2M traffic is dominated by UL traffic as
the M2M devices sense and monitor and thereby generate the data to send
\cite{shafiq2013large}. M2M traffic will represent an important segment in the upcoming
5G wireless systems. One of the new features in 5G systems is the possibility to offer
ultra-reliable connections. This will bring a new quality in the support of M2M traffic,
as services can be built under the assumption that the M2M device will be able to deliver
its data with very high reliability.

There are multiple ways to improve the reliability of UL transmissions, such as using
higher transmission power or antenna diversity. On the other hand, the trend of dense
deployment~\cite{osseiran2013foundation} of Small-cell Base Stations (SBSs) in
heterogeneous networks brings the infrastructure close to the terminals and brings the
possibility to improve the UL transmission by careful cell association. The
works~\cite{elshaer2014downlink,smiljkovikj2014analysis} show that, due to the difference
in the DL/UL power, it is beneficial to DL/UL decoupled access (DUDe), such that the
terminal receives from one BS, but transmits to another one. DUDe can be understood as
the use of selection macro-diversity.

\begin{figure}[tb]
\centering
\includegraphics[width=0.5\columnwidth]{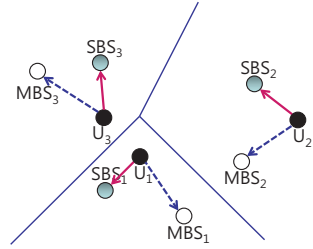}
\caption{Example of double association. Each uplink device associates with one MBS and
one SBS. } \label{F:double_assoc_example}
\end{figure}

In this letter we use the fact that multiple densely deployed SBSs can be in the
proximity of the terminal, such that the terminal can be simultaneously associated with
two or more BSs in the UL and its UL packets are transmitted to all of them.
Specifically, we treat the basic variant of \emph{double association}, depicted on
Fig.~\ref{F:double_assoc_example}, in which a terminal is associated with one Macro-cell
BS (MBS) and one SBS. Clearly, the UL broadcast improves the reliability compared to the
DUDe and single-point UL association. Note that dual connectivity has already been
considered in the release 12 of LTE~\cite{astely2013lte} with the main purpose to improve
downlink connectivity; however, each individual connection is put at a different
frequency, i.e. a single transmission is not received by more than one access points. In
\cite{falconetti2011uplink}, the authors also consider the joint reception of the uplink
signal where the uplink transmission of a user is received by more than one node.
However, the results are largely based on simulations.

The results in this letter show that double association can lead to significant improvement in
reliability, expressed through the probability of successful reception. The performance
of the double association is analyzed by using stochastic geometry \cite{Stoyan1995} for
the distribution of the terminals, MBSs and SBSs.

\section{System Model}

We treat the scenario of a wireless heterogeneous network, composed of MBSs and SBSs,
where each active user has an uplink (UL) transmission. The downlink (DL) transmission
takes place in a time-frequency resource that is orthogonal to the one used for UL.

We assume that the MBSs and the SBSs are randomly located according to the 2-dimensional
homogeneous Poisson point processes (PPP) $\Phi$ and $\Phi_s$ with node densities
$\lambda$ and $\lambda_s$, respectively. The connection between the user and the MBS is
assumed to use OFDMA-like orthogonal multiple access scheme, such that there is only one
user per MBS using the same resource. To model this, the total area is considered as a
Voronoi diagram, with the MBSs being the seeds of the diagram, such that each MBS defines
one Voronoi cell. In each Voronoi cell one active UL user is deployed in a uniform random
way across the cell area. Hence, the density of active UL users is also $\lambda$;
however, the users are not generated by a Poisson point process, unlike the MBSs and
SBSs. The distribution of users is a Poisson-Voronoi perturbed lattice
\cite{blaszczyszyn2015clustering} which is hard to analyze, but in order to make our
analysis tractable, we will approximate it with a PPP with intensity $\lambda$. The
tractability and accuracy of this approach has been demonstrated in
\cite{novlan2013analytical}. The \emph{typical MBS} is placed at the origin and the
nearest user to the origin is the associated user. The nearest SBS to the user associated
with the typical MBS is considered as the \emph{typical SBS}. All performance figures are
calculated for these two typical access points.

The double association is specified as follows. The $i-$th user associates with the MBS
$M(i)$ deployed in the $i-$th Voronoi cell. Additionally, the $i-$th user's transmission
can reach the SBS $S(i)$ from which it receives the highest average power, i.e. the SBS
that is closest geometrically:
\begin{equation}\label{E:da}
{S(i)} = \arg \mathop {\min }\limits_{j \in {\Phi _s}} r_{i,j},
\end{equation}
where $r_{i,j}$ is the distance between the $i-$th active user and $j-$th SBS. The MBS
acts as a primary access point, but the user transmission is simultaneously received by
the MBS and the SBS. This scheme can be generalized to $k+1$ associations by having $k$
additional associations to the $k$ nearest SBSs, i.e. each user transmission has $k$
additional degrees of diversity.

All UL transmissions use the same spectrum and interfere with each other. The effects of
path loss and fading are compensated by fractional power control (FPC)
\cite{castellanos2008performance,novlan2013analytical}. The received signal strength at
$B(i)$ from the $i-$th active user is:
\begin{equation}\label{eq:FPC}
G_{i,B(i)}r_{i,B(i)}^{-\alpha}\min(r_{i,M(i)}^{\alpha\epsilon}P,\bar{P}),
\end{equation}
where $r_{i,B(i)}$ is the distance between the $i-$th active user and $B(i) \in \{M(i),
S(i) \}$, $\alpha$ is the path-loss coefficient and $G_{i,B(i)}$ is an exponential random
variable with unit mean in order to model the Rayleigh fading on the link. The time is
slotted and the channel gain $G_{i,B(i)}$ is invariant in a slot. The transmit power is
the minimum value between the maximum transmit power level constraint $\bar{P}$ and the
power control component $r_{i,M(i)}^{\alpha\epsilon}P$, where $P$ is default transmit
power level and $\epsilon$ is a power control factor (PCF) that can have values
$0\!\leq\!\epsilon\!\leq\!1$. For $\epsilon\!=\!0$, no power control is applied, while
$\epsilon\!=\!1$ corresponds to full channel inversion. The use of $r_{i,M(i)}$ indicates
that  power control is applied with respect to the distance to the associated MBS. Due to
the dense deployment, we assume that the network is interference-limited and the effect
of the noise is ignored. The success of the transmission is determined by the received
signal-to-interference ratio (SIR) $\gamma_{B(i)}$ of the associated base station $B(i)
\in \{M(i), S(i) \}$, given by:
\begin{equation}\label{E:isir}
{\gamma _{B(i)}} \!\!=\!\!
\frac{{{G_{i,B(i)}}r_{i,B(i)}^{ - \alpha
}\!\min(r_{i,M(i)}^{\alpha\epsilon},\!\bar{P})}}{{\sum\limits_{k \ne i}\!\!
{{G_{k,B(i)}}r_{k,B(i)}^{ - \alpha }\!\min(r_{k,M(k)}^{\alpha\epsilon},\!\bar{P})}}}
\!\!=\!\! \frac{{{G_{i,B(i)}}r_{i,B(i)}^{ - \alpha
}\!\min(r_{i,M(i)}^{\alpha\epsilon},\!\bar{P})}}{I},
\end{equation}
where $I$ denotes the aggregate interference. For a given target SIR threshold $\beta$,
the transmission is successful if ${\gamma _{B(i)}} \geq \beta$. The data rate is a
function of the target SIR following Shannon's formula $\log_2\left(1\!+\!\beta\right)$
with a unit bandwidth. The transmit powers of the interfering users
($\min(r_{k,M(k)}^{\alpha\epsilon},\!\bar{P})$ in \eqref{E:isir}) are actually not
independent, since the area of the Voronoi cells of the adjacent users are correlated. In
order to the analytical tractability, we will assume that the transmit powers of the
interfering users are independent. The similar assumption is also made in
\cite{novlan2013analytical}.

As a reference scheme, the user $i$ associates with the one of the BSs (MBS or SBS),
denoted by  $j^*$, which has the maximal average received signal strength in the UL:
\begin{equation}\label{E:de}
{j^*} = \arg \mathop {\max }\limits_{j \in {\Phi _b} \cup {\Phi _s}} r_{i,j}^{ - \alpha }P.
\end{equation}
Note that the ``classical'' association is based on the received DL signal. Our reference
scheme is thus related to the newly proposed DL/UL Decoupling (DUDe)
\cite{elshaer2014downlink}, where the UL association is done according to (\ref{E:de})
and is decoupled from the DL association. We call this scheme as a single association
(SA). We use $p_s^{DA}$ to denote the success probability of DA. In the following
sections, we analyze the success probability.

\section{Reliability Analysis of Double Association Scheme}
\label{sec:RelAnalysis}

Now we consider the success probability $p_s^{DA}$ of double association. The user
connects to both the MBS and the SBS. Then the success probability is:
\begin{align}\label{E:psDA1}
  {p_s^{DA}} = 1 - \Pr \left[ {{\gamma _{M}} < \beta ,{\gamma _{S}} < \beta } \right],
\end{align}
where $\gamma_{M}$ and $\gamma_{S}$  denote the received SIR at the typical MBS and SBS.
Eq.~\eqref{E:psDA1} means the transmission is only failed if both MBS and SBS cannot
receive the data. In our system model, MBSs and SBSs are deployed by
independent point processes, such that these two probabilities are independent. We can
rewrite \eqref{E:psDA1} as follows:
\begin{align}\label{E:psDA2}
  {p_s^{DA}} = 1 - \left( {1 - \Pr \left[ {{\gamma _{M}} \geq \beta } \right]} \right)\left( {1 - \Pr \left[ {{\gamma _{S}} \geq \beta } \right]} \right).
\end{align}
The probability $\Pr \left[ {{\gamma _{M}} \geq \beta } \right]$ can be expressed as:
\begin{align}\label{E:psmbs}
\Pr \left[ {{\gamma _{M}} \geq \beta } \right] \!&=\! \mathbb{E}_r\left[\Pr \left[ {{\gamma _{M}} \geq \beta } |r \right]\right] \nonumber \\
\!&=\! \int_0^\infty  \!\!\!{\Pr \left[ {\frac{{G{r^{ - \alpha }}{\min \left( {{r^{\alpha
\epsilon }}P,\bar P} \right)}}}{{{I_M}}} \!\geq\! \beta } \right]{f_R}\left( r
\right)\!dr},
\end{align}
where $r$ denotes the distance to typical MBS. The FPC is performed based on the distance
to the associated MBS. The term $I_M$ denotes the interference at typical MBS. The
communication distance $r$ is a random variable with probability density function
$f_R(r)$. As stated already, we approximate the user deployment process as PPP with
intensity $\lambda$, such that the distance distribution becomes $f_R(r)={2\pi \lambda
r{e^{ - \pi \lambda {r^2}}}}$. The communication distances, which are the distances of
the UL users to their associated MBSs, might be identically distributed, but not
independent. The dependence is caused by the restriction that only one UL user can be
situated in each Voronoi cell. This implies that, on average, the communication distances
are closer compared to the communication distance in PPP since there is no restriction in
PPP (See Fig.~2 in \cite{novlan2013analytical}). Hence, the success probability in PPP
offers a lower bound to the original success probability,  \eqref{E:psmbs} expressed as
follows:
\begin{align}\label{E:psmbs2}
\!&\geq\!
 \int_0^\infty  {\Pr \left[ {\frac{{G{r^{ - \alpha }}\min \left( {{r^{\alpha \epsilon }}P,\bar P} \right)}}{{{I_M}}} \ge \beta } \right]2\pi \lambda r{e^{ - \pi \lambda {r^2}}}dr} \nonumber \\
 &\stackrel{(a)}= \int_0^\infty  {{\mathbb{E}_{{I_M}}}\left[ {\exp \left( { - \beta {r^\alpha }\min {{\left( {{r^{\alpha \epsilon }}P,\bar P} \right)}^{ - 1}}{I_M}} \right)} \right]2\pi \lambda r{e^{ - \pi \lambda {r^2}}}dr} \nonumber \\
 &\stackrel{(b)}= \int_0^{{{\hat P}^{\frac{1}{{\alpha \epsilon }}}}} {{\mathbb{E}_{{I_M}}}\left[ {\exp \left( { - \frac{{\beta {r^\alpha }{r^{ - \alpha \epsilon }}}}{P}{I_M}} \right)} \right]2\pi \lambda r{e^{ - \pi \lambda {r^2}}}dr} \nonumber \\
 &\qquad\qquad+\int_{{{\hat P}^{\frac{1}{{\alpha \epsilon }}}}}^\infty  {{\mathbb{E}_{{I_M}}}\left[ {\exp \left( { - \frac{{\beta {r^\alpha }}}{{\bar P}}{I_M}} \right)} \right]2\pi \lambda r{e^{ - \pi \lambda {r^2}}}dr} \nonumber \\
 &\stackrel{(c)}= \!\!\int_0^{{{\hat P}^{\frac{1}{{\alpha \epsilon }}}}} \!\! {{{\cal L}_{{I_M}}}\left( {{s_1}} \right)2\pi \lambda r{e^{ - \pi \lambda {r^2}}}dr}  \!+\! \int_{{{\hat P}^{\frac{1}{{\alpha \epsilon }}}}}^\infty \!\! {{{\cal L}_{{I_M}}}\left( {{s_2}} \right)2\pi \lambda r{e^{ - \pi \lambda {r^2}}}dr}
\end{align}
where (a) comes from the property of exponential random variable and takes the
expectation of $I_M$. In (b), the integration is split into two parts, as the transmit power
is varied with $r$ and $\bar{P}$ is selected when $r>{\hat P}^{\frac{1}{\alpha
\epsilon}}$, where $\hat P = \bar P/P$. For (c), let us define ${s_1} = {\beta {r^\alpha
}{r^{ - \alpha \epsilon }}}/P$ and ${s_2} = {\beta {r^\alpha }}/{\bar P}$, then the
expectation of $I_M$ becomes Laplace functional as follows:
\begin{align}\label{E:laplacembs1}
{{\mathcal{L}}_{{I_M}}}\left( {{s_1}} \right) &= \exp \Bigg(  { - 2\pi \lambda \int_r^\infty  {\frac{{\beta {r^{\left( {1 - \epsilon } \right)\alpha }}{v^{ - \alpha }}\hat P}}{{1 + \beta {r^{\left( {1 - \epsilon } \right)\alpha }}{v^{ - \alpha }}\hat P}}{e^{ - \pi \lambda {{\hat P}^{\frac{2}{{\alpha \epsilon }}}}}}vdv} } - \nonumber \\
2\pi &\lambda \!\!\! \int_r^\infty  \!\!\!\!\! {\int_0^{{{\hat P}^{\frac{1}{{\alpha \epsilon }}}}} \!\!\!\!\!\!\! {\frac{{\beta {r^{\left( {1 - \epsilon } \right)\alpha }}{v^{ - \alpha }}{x^{\alpha \epsilon }}}}{{1 \!\!+\! \beta {r^{\left( {1 - \epsilon } \right)\alpha }}{v^{ - \alpha }}{x^{\alpha \epsilon }}}}2\pi \lambda x\exp \left( { - \pi \lambda {x^2}} \right)\!vdxdv} }    \Bigg)
\end{align}
\begin{align}\label{E:laplacembs2}
{{\mathcal{L}}_{{I_M}}}\left( {{s_2}} \right) &= \exp \Bigg(  { - 2\pi \lambda \int_r^\infty  {\frac{{\beta {r^\alpha }{v^{ - \alpha }}}}{{1 + \beta {r^\alpha }{v^{ - \alpha }}}}{e^{ - \pi \lambda {{\hat P}^{\frac{2}{{\alpha \epsilon }}}}}}vdv} } - \nonumber \\
2\pi &\lambda \!\!\! \int_r^\infty  \!\!\!\!\! {\int_0^{{{\hat P}^{\frac{1}{{\alpha \epsilon }}}}} \!\!\!\!\!\!\! {\frac{{\beta {r^\alpha }{v^{ - \alpha }}{x^{\alpha \epsilon }}/\hat P}}{{1 \!+\! \beta {r^\alpha }{v^{ - \alpha }}{x^{\alpha \epsilon }}/\hat P}}2\pi \lambda x\!\exp \! \left( { - \pi \lambda {x^2}} \right)\!vdx} dv}
\Bigg).
\end{align}
The derivation of \eqref{E:laplacembs1} and \eqref{E:laplacembs2} uses the property of
PPP \cite{Stoyan1995} and the integration of $v$ is made by using the distance from the
interferers as a variable. The integration starts from $r$ due to the property of the
Voronoi cells and no interferer can be closer to the typical MBS than the typical user
associated with that MBS. The integration of $x$ uses the independent assumption of
transmission powers of interferers and the integration range is limited to ${\hat
P}^{\frac{1}{{\alpha \epsilon}}}$ for the same reason as \eqref{E:psmbs2}-(b).

The probability ${\Pr \left[ {{\gamma _{S}} \geq \beta } \right]}$ can be obtained in a
similar manner:
\begin{align}\label{E:pssbs}
\Pr \left[ {{\gamma _{S}} \geq \beta } \right] &\!\simeq\!
\int_0^\infty  \Bigg\{ \int_0^{{{\hat P}^{\frac{1}{{\alpha \epsilon }}}}} {{{\cal L}_{{I_s}}}\left( {{s_1}} \right)2\pi \lambda r{e^{ - \pi \lambda {r^2}}}dr} + \nonumber \\
&\qquad {{\cal L}_{{I_s}}}\left( {{s_2}} \right){e^{ - \pi \lambda {{\hat P}^{\frac{2}{{\alpha \epsilon }}}}}} \Bigg\}2\pi {\lambda _s}y{e^{ - \pi {\lambda _s}{y^2}}}dy,
\end{align}
where ${s_1} = \frac{{\beta {y^\alpha }{r^{ - \alpha \epsilon }}}}{P}$, ${s_2} =
\frac{{\beta {y^\alpha }}}{{\bar P}}$, and $I_S$ denotes the interference at typical SBS.
The Laplace functionals are
\begin{align}\label{E:laplacesbs1}
{{\mathcal{L}}_{{I_S}}}\left( {{s_1}} \right) &= \exp \Bigg( { - 2\pi \lambda \int_0^\infty  {\frac{{\beta {y^\alpha }{r^{ - \alpha \epsilon }}{u^{ - \alpha }}\hat P}}{{1 + \beta {y^\alpha }{r^{ - \alpha \epsilon }}{u^{ - \alpha }}\hat P}}{e^{ - \pi \lambda {{\hat P}^{\frac{2}{{\alpha \epsilon }}}}}}udu}  - } \nonumber \\
2\pi \lambda & \!\!\! \int_0^\infty  \!\!\!\!\! {\int_0^{{{\hat P}^{\frac{1}{{\alpha \epsilon }}}}} \!\!\!\!\!\!\! {\frac{{\beta {y^\alpha }{r^{ - \alpha \epsilon }}{u^{ - \alpha }}{x^{\alpha \epsilon }}}}{{1 \!+\! \beta {y^\alpha }{r^{ - \alpha \epsilon }}{u^{ - \alpha }}{x^{\alpha \epsilon }}}}\!2\pi \lambda x\!\exp \!\left( { - \pi \lambda {x^2}} \right)\!udxdu} }  \Bigg),
\end{align}
\begin{align}\label{E:laplacesbs2}
{{\mathcal{L}}_{{I_S}}}\left( {{s_2}} \right) &=\! \exp \Bigg( { - 2\pi \lambda \int_0^\infty  {\frac{{\beta {y^\alpha }{u^{ - \alpha }}}}{{1 + \beta {y^\alpha }{u^{ - \alpha }}}}{e^{ - \pi \lambda {{\hat P}^{\frac{2}{{\alpha \epsilon }}}}}}udu} } - \nonumber \\
2\pi \lambda  \!\!\! \int_0^\infty  &\!\!\!\!\! {\int_0^{{{\hat P}^{\frac{1}{{\alpha \epsilon }}}}} \!\!\!\!\!\!\!\! {\frac{{\beta {y^\alpha }{u^{ - \alpha }}{x^{\alpha \epsilon }}/\hat P}}{{1 + \beta {y^\alpha }{u^{ - \alpha }}{x^{\alpha \epsilon }}/\hat P}}2\pi \lambda x\exp \left( { - \pi \lambda {x^2}} \right)udxdu} }  \Bigg).
\end{align}
For the SBSs, the distance of the interferers begins at zero since the interferer could
be closer than the typical user. Using \eqref{E:psmbs2}-\eqref{E:laplacesbs2}, the
success probability \eqref{E:psDA2} can be computed. Even though $p_s^{DA}$ is not a
closed form, it can be easily computed numerically. With specific parameters
$\alpha\!=\!4$ and $\epsilon\!=\!0$, a lower bound on $p_s^{DA}$ can be expressed in a
closed form:
\begin{align}\label{E:psDAspecial}
  {p_s^{DA}} \!\geq\! 1 \!-\! \left( {1 - \frac{1}{{1 + \sqrt \beta  \arctan \left( {\sqrt \beta  } \right)}}} \right)\left( {1 - \frac{{2{\lambda _s}}}{{\pi \sqrt \beta  \lambda  + 2{\lambda _s}}}} \right).
\end{align}

\section{Performance Evaluation}

We use Monte Carlo simulation in order to assess the performance of the different
association schemes and verify the analytical derivations. The node densities of MBSs and
active uplink devices are set to $\lambda=0.01$, while the node density of SBSs is varied
from $\lambda_s=0.02$ to $\lambda_s=0.05$. The path-loss exponent is $\alpha=4$ and the
default transmit power for UL user is $P=30$dBm. The different PCF values $\epsilon=0,
0.5, 1$ and target thresholds $\beta=$0dB, 5dB are used.

\begin{figure}[tb]
\centering
\includegraphics[width=0.9\columnwidth]{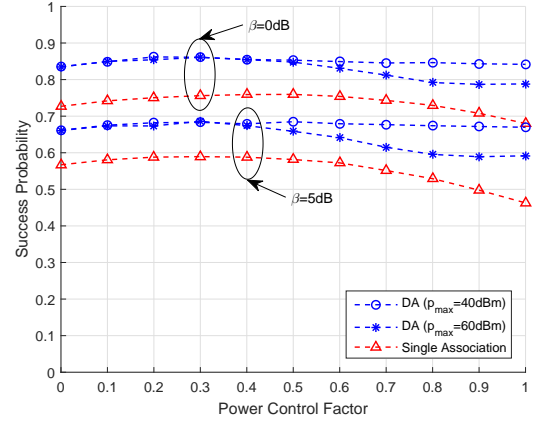}
\caption{Success probability as a function of power control factor ($\lambda\!=\!0.01$, $\lambda_s\!=\!0.02$). }
\label{F:sprob_vs_power_control_factor}
\end{figure}

The transmission reliability is evaluated through the probability of success.
Fig.~\ref{F:sprob_vs_power_control_factor} shows success probability as a function of
power control factor ($\epsilon$). The reliability of DA is distinctly improved compared
to that of SA. The lower target SINR threshold leads to a higher reliability performance,
while, as expected, DA always outperforms SA. The small maximum power constraint (40dBm)
will reduce the interference power and increase the reliability of DA compared to a
higher constraint (60dBm). Even though utilizing lower PCF will increase the reliability,
the effect of FPC is insignificant for DA. On the other hands, it is important to choose
proper PCF for SA, as it is more sensitive and results in performance variation.

\begin{figure}[tb]
\centering
\includegraphics[width=0.9\columnwidth]{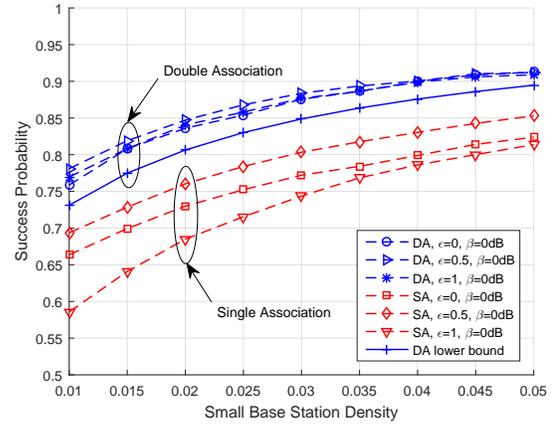}
\caption{Success probability as a function of the small-cell base station density ($\lambda\!=\!0.01$, $\beta\!=\!0$~dB, $\bar{p}\!=\!50$~dBm).}
\label{F:sprob_vs_sbs_0db}
\end{figure}

\begin{figure}[tb]
\centering
\includegraphics[width=0.9\columnwidth]{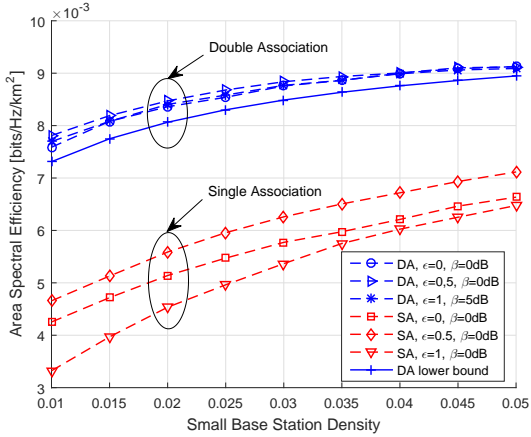}
\caption{Area spectral efficiency as a function of the small-cell base station density ($\lambda\!=\!0.01$, $\beta\!=\!0$~dB, $\bar{p}\!=\!50$~dBm).}
\label{F:ase_vs_sbs_0db}
\end{figure}

Fig.~\ref{F:sprob_vs_sbs_0db} shows the success probability as a function of the node
density of SBSs. Already SA increases the success probability, but this is further
increased by a double association. The analytical result shows a tight lower bound of the
performance.

In order to assess the network throughput performance, we evaluate the \emph{area
spectral efficiency}, the product of the successfully transmitting user density and the
data rate per node. Fig.~\ref{F:ase_vs_sbs_0db} illustrates the area spectral efficiency
as a function of the node density of SBSs. For the SBSs, multiple uplink users are
connected to the same SBSs. Since the success of transmission is determined by the
received SIR and the target threshold, only one user can succeed if the target threshold
is $\beta \geq 1$.  It can be seen that DA is superior to SA also in terms of area
spectral efficiency.

\begin{figure}[tb]
\centering
\includegraphics[width=0.9\columnwidth]{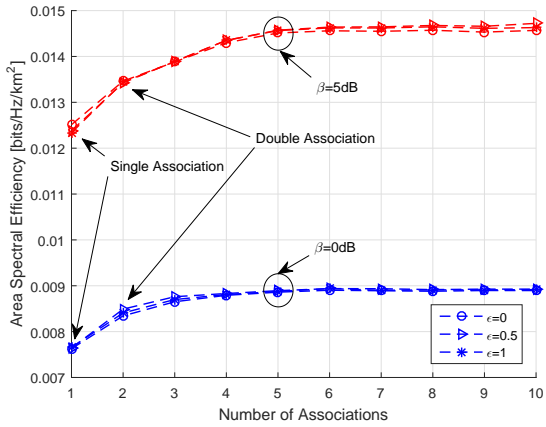}
\caption{Area spectral efficiency as a function of the number of associations ($\lambda\!=\!0.01$, $\lambda_s\!=\!0.02$, $\beta\!=\!0$~dB, $\bar{p}\!=\!50$~dBm).}
\label{F:ase_vs_assoc}
\end{figure}

The derivation of the probabilities for $k\!+\!1$ associations is much more involved
compared to the special case $k\!=\!1$ and requires elaboration that is outside the scope
of this letter.\footnote{To quantify the performance, it is needed to calculate the
distance distribution of $n-$th nearest SBSs. For the purpose the approaches proposed in
\cite{nigam2014coordinated,baccelli2015stochastic} can be used.} Here we use simulation
results to show how the reliability depends on $k$. By increasing the number of
associations, the reliability of the transmission is improved, but rather saturated as
the number of associations increase. The area spectral efficiency performance is depicted
on Fig.~\ref{F:ase_vs_assoc}. Increasing the number of associations is more favorable
with as the target threshold increases. For example, having five associations is still
beneficial for $\beta\!=\!5$dB, however, in case of $\beta\!=\!0$dB, the performance is
almost saturated with three associations. Nevertheless, it can be seen clearly that the
major performance increase comes when moving from single- to a double association.

\vspace{3mm}

\section{Concluding Remarks}
We have considered methods for network association that improve the reliability of uplink
transmissions in dense wireless heterogeneous networks. We have used stochastic geometry
in order to analyze the performance of the schemes. The extensive simulation results show
that double association, to one macro-cell Base Station (BS) and one small-cell BS,
remarkably improves the probability of successful uplink transmission. As for the next
steps, it is interesting to investigate the performance of a more complex method of
processing the received data jointly across base stations. Finding the proper power
control strategy is also interesting topic. If Successive Interference Cancellation (SIC)
is applied along with double association, some of the uplink users can reduce the
transmission power in order to catalyze the SIC process.

\end{document}